\documentclass{ws-p8-50x6-00}

\def\ifmath#1{\relax\ifmmode #1\else $#1$\fi}%
\def\rd{\ifmath{{\mathrm{d}}}}
\def\rL{\ifmath{{\mathrm{L}}}}

\def\final{\ifmath{{\mathrm{final}}}}
\def\incl{\ifmath{{\mathrm{incl}}}}
\def\max{\ifmath{{\mathrm{max}}}}

\def\jet{\ifmath{{\mathrm{jet}}}}
\def\naive{\ifmath{{\mathrm{naive}}}}

\begin{document}

\title{Transverse Energy and Minijets in High Energy Collisions}

\author{G\"osta Gustafson}

\address{Dept. of Theor. Physics, Lund University, 
S\"olvegatan 14A, 22362 Lund, Sweden}


\maketitle

\abstracts{
Minijet production and transverse energy are important not only to
understand hadronic collisions, but also for the interpretation of
nucleus collisions at RHIC and LHC, where it determines the ``initial
conditions'' for the flow in a hadronic soup or a plasma. For high
collision energies and small $q_\perp$ (minijets) we enter the BFKL
region. This implies that we must take into account off-shell parton
cross sections and non-integrated structure functions
($k_\perp$-factorization). It is also essential to avoid double
counting, as one emitted parton is a participant in two different
subcollisions. The LDC model, developed in Lund to describe DIS,
provides a very convenient formalism to handle these problems. The
result is a dynamical suppression of minijets for small $q_\perp$. The
resulting $E_\perp$-flow is similar to the result from a ''naive''
calculation based on integrated structure functions with a $q_\perp$
cut-off around 2 GeV.}
\section{Introduction}
In hadronic collisions, jets with {\em large} $q_\perp$ can be described by 
the expression
\begin{equation}
\frac{\rd\sigma}{\rd q_\perp^2} \sim F(x_a,q_\perp^2)
F(x_b,q_\perp^2) \frac{\alpha_s^2(q_\perp^2)}{q_\perp^4}\ .
\label{naive}
\end{equation}
The structure functions $F(x,q_\perp^2)$ can here be described by
DGLAP evolution; the initial parton cascades are ordered in $q_\perp$
towards the hard subcollision. The cross section for quasireal partons
is approximately $\sim \alpha_s^2(q_\perp^2) / q_\perp^4 $. For
smaller $q_\perp$, minijets, we enter the BFKL domain, with important
contributions from non-ordered parton chains. The expression in
eq.~(\ref{naive}) blows up for small $ q_\perp $, and in
phenomenological applications a cutoff around 2 GeV is needed to
reproduce experimental data.\cite{cutoff} This phenomenological
cutoff appears to grow with energy, which makes it difficult to make
predictions for very high energies, e.g. at the LHC. Thus we have two
essential problems:

1. For non-ordered chains we get contributions from scattering of
off-shell partons with $k_{\perp a}$ and/or $k_{\perp b} > q_\perp$,
for which the cross section has to be modified. From $
k_\perp$-factorization, this problem can be handled by non-integrated
structure functions ${\cal F}(x,k_\perp^2)$ and off-shell
parton-parton cross sections.

2. In each chain as in fig.~\ref{fig:fans} there are many final state
partons (minijets). Each final parton is connected to two links, which
implies that it is a component in {\em two} different
subcollisions. Therefore, we must be cautious to avoid double counting.

Both these problems can be conveniently treated within the Linked
Dipole Chain (LDC) model.\cite{LDC,LDCMC} This model interpolates
between DGLAP and BFKL. In a single formalism it describes different
types of reactions in DIS: ``Normal DIS'', boson-gluon fusion, and
hard resolved $\gamma$p scattering. When applied to hh or AA
collisions it implies that the inclusive jet cross section is reduced
for smaller $q_\perp$ {\em cf}. to the expression in
eq.~(\ref{naive}). The total $E_\perp$ flow is similar to the
``naive'' result if this has a low $q_\perp$ cut-off $\sim 2$ GeV.

\section{DIS} 
\begin{figure}[t]
\mbox{
\epsfxsize=8pc 
\epsfbox{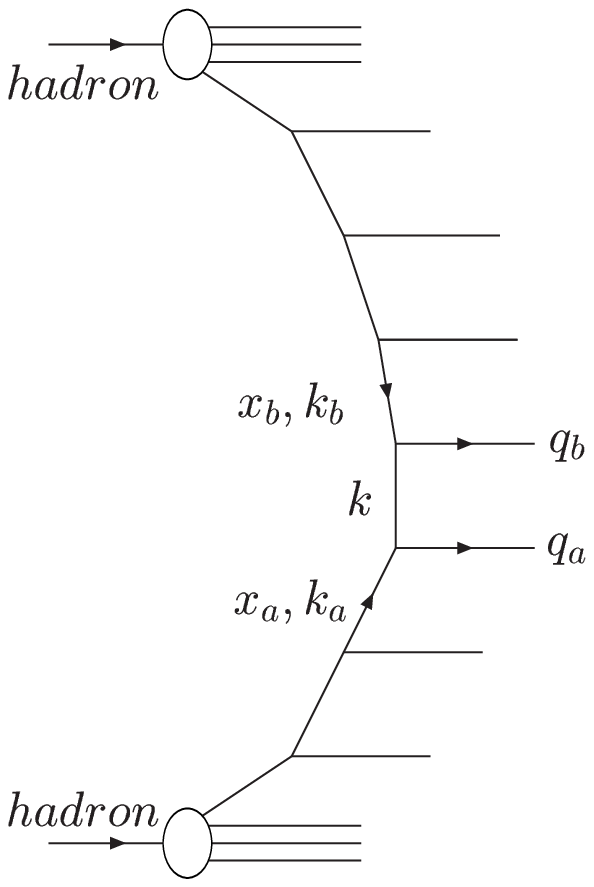}} 
\mbox{
\epsfxsize=20pc 
\epsfbox{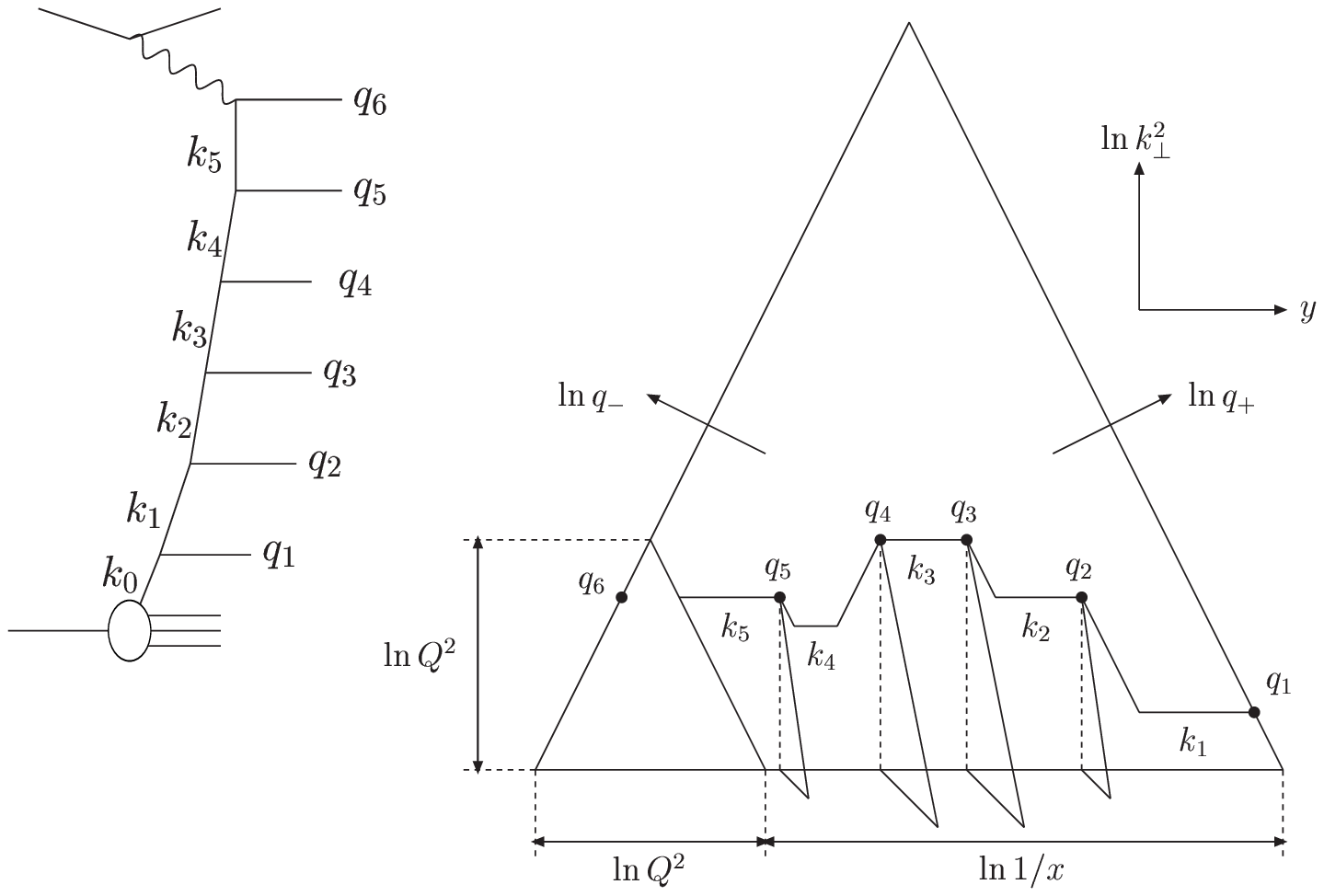}} 
\caption{ Fan diagrams in hh collisions and in DIS. In the LDC model
the initial state emissions $q_i$ form a chain in the
$(y,\kappa=\ln(k_{\perp}^2))$-plane. Final state radiation is allowed
in the region below the horizontal lines. \label{fig:fans}}
\end{figure}

When $Q^2$ is large and $x$ not too small (in the DGLAP region) the
dominant contribution to $F_2$ is given by gluon chains, which are
ordered in $x$ (or $q_\rL$) and in $k_\perp$. Each chain as in
fig.~\ref{fig:fans} gives a contribution
\begin{equation}
\prod_i^n \bar{\alpha} \frac{\rd x_i}{x_i} 
\frac{\rd q_{\perp,i}^2}{q_{\perp,i}^2} \,\,\,\,
{\mathrm {where}} \,\,\,\, \bar{\alpha} \equiv \frac{3\alpha_s}
{\pi} \equiv \frac{\alpha_0}{\ln(q_\perp^2)}.
\label{vikt}
\end{equation}
Summing over different values of $n$ gives the double leading log
result $F \sim \exp (2\sqrt{\alpha_0 \ln 1/x \ln \ln Q^2})$. The
gluons in such a chain constitute the {\em initial state} radiation.
To obtain the properties of the complete final state, we must add {\em
final state} radiation within angular ordered regions.

When $Q^2$ is moderate and $x$ small (in the BFKL region), the
$k_\perp$-ordered region is small, and non-ordered contributions are
important although suppressed. The result is a power-like increase of
$F$ for small $x$, $F \sim x^{-\lambda}$.

In the {\em interpolation region} we must calculate suppressed
contributions from non-ordered chains. It is then necessary to specify
the separation between initial state radiation and final state
radiation. This is not given by Nature; it has to be defined by the
calculation scheme. A particular scheme was chosen by Ciafaloni,
Catani, Fiorani, and Marchesini (the CCFM model).\cite{CCFM} Here the
initial state radiation is ordered in {\em angle} (or rapidity) and
{\em energy} (or $q_+=q_0 + q_\rL$). The contribution from each chain is
then determined by specific non-Sudakov form factors.

\subsection{The Linked Dipole Chain model}
The Linked Dipole Chain (LDC) model \cite{LDC,LDCMC} is a
reformulation and generalization of the CCFM result in a scheme, where
more gluons are treated as final state radiation. The initial chain is
ordered in $q_+=q_0 + q_\rL$ {\em and} $q_-=q_0 - q_\rL$, and $q_\perp$
satisfies $q_\perp > min(k_{\perp,i},k_{\perp,i-1})$. This implies
that there are fewer chains. One LDC chain corresponds to a set of
CCFM chains. It then turns out that all the corresponding non-Sudakov
form factors add up to just unity. Thus the contribution from each
such chain is given by an expression identical to eq.~(\ref{vikt}). We
note in particular that this expression is totally left-right
symmetric, meaning that we get the same result if we start the chain
in the photon end, instead of in the proton end.

We can express this result in the link momenta $k_i$, instead of the
final state momenta $q_i$.  Using the relations $\rd^2 q_{\perp,i} = \rd^2
k_{\perp,i}$ and $q_{\perp,i}^2 \approx \max( k_{\perp,i}^2,
k_{\perp,i-1}^2)$ we find the following weights
\begin{equation} 
\frac{\rd^2 q_{\perp,i}}{q_{\perp,i}^2} \approx \frac{\rd^2 k_{\perp,i}}
{k_{\perp,i}^2}
\,\,{\mathrm {for}}\,\, k_{\perp,i} > k_{\perp,i-1};\,\,
\frac{\rd^2 q_{\perp,i}}{q_{\perp,i}^2} \approx \frac{\rd^2 k_{\perp,i}}
{k_{\perp,i}^2} 
\cdot \frac{ k_{\perp,i}^2}{k_{\perp,i-1}^2}\,\,{\mathrm {for}}\,\,
k_{\perp,i} < k_{\perp,i-1}.
\label{step-up-down}
\end{equation}
Thus, for a step down in $k_\perp$ we have an extra suppression factor 
$ k_{\perp,i}^2/k_{\perp,i-1}^2$. This implies that if the chain 
goes up to $k_{\perp,\max}$ and then down to $k_{\perp,\final}$ we obtain a 
factor $1/k_{\perp,\max}^4$, which corresponds to the cross section for a 
hard parton-parton subcollision.
An important feature is that different types of reactions can be
described in the same formalism. Thus ``normal'' DIS corresponds to
the case when $k_{\perp,\final}^2 < Q^2$, boson-gluon fusion to
$k_{\perp,\final}^2 > Q^2$, while resolved $\gamma$p scattering is
obtained when $k_{\perp,\max}^2 > k_{\perp,\final}^2 (> Q^2)$.

MC programs have been developed for both the CCFM model and the LDC
model and compared to HERA data.\cite{HJLL} Programs for the two
models can fit $F_2$ and essential properties of the final states. A
still puzzling feature is that both programs can fit the forward jets
only provided the nonsingular terms in the splitting functions are
suppressed.

A running coupling $\alpha_s$ favours small $k_\perp$ in the
chain. This implies that the structure function factorizes for small
$x$-values.\cite{LDC} For the results in the next section we use the
expression $\mathcal{F} \sim x^{-0.3} (\ln k_\perp^2)^2$ which is
motivated by the MC fit to HERA data.\cite{LDCMC}
\section{Jet Cross Section in Hadronic Collisions}
The results presented in this section are obtained in collaboration
with Gabriela Miu.\cite{miu} For a resolved $\gamma$p scattering the
symmetry of the formalism implies that it can be interpreted as
evolution from both ends towards a central hard scattering. This
obviously works also for pp or AA scattering. Study a single link
with notation as in fig.~\ref{fig:fans}. There are three different
possibilities:

1. $k_\perp > k_{\perp,a},k_{\perp,b} \Rightarrow q_{\perp,a} \approx
q_{\perp,b} \approx k_\perp$ which has a weight $ k_\perp^{-4}$.

2. $k_{\perp,a} > k_{\perp},k_{\perp,b} \Rightarrow q_{\perp,a}
\approx k_{\perp,a}$, $q_{\perp,b} \approx k_\perp$, with weight
$k_{\perp,a}^{-2} \cdot k_\perp^{-2}$.

3. $k_\perp < k_{\perp,a},k_{\perp,b} \Rightarrow q_{\perp,a} \approx
k_{\perp,a}$, $q_{\perp,b} \approx k_{\perp,b}$, with weight
$k_{\perp,a}^{-2} \cdot k_{\perp,b}^{-2}$.
 
In each case we have also a factor $\alpha_s(q_{\perp,a}^2) \cdot
\alpha_s(q_{\perp,b}^2)$.  We note that there is a suppression when
the virtuality of the colliding particles ($k_{\perp,a}^2$ and
$k_{\perp,b}^2$) are larger than the momentum transfer
$k_\perp^2$. This implies that $\sigma_{\incl}$ does not blow up for
small $q_\perp$-values.  The inclusive cross section can be expressed
in terms of the non-integrated structure functions
$\mathcal{F}(x,k_\perp^2)$.
\begin{equation}
\frac{\rd \sigma_{\incl}}{\rd q_\perp^2 \rd y} \sim 
\int \mathcal{F}(x_a, k_{\perp a}^2) \cdot \mathcal{F}(x_b, 
k_{\perp b}^2) \nonumber \\ 
\cdot \frac{1}{2} \cdot \frac{\rd \hat\sigma}{\rd q_\perp^2} 
(q_\perp^2,  k_{\perp a}^2, k_{\perp b}^2, \hat s)\ ,
\label{incljetdistr} 
\end{equation} 
where $\hat s = x_a x_b s$ and $y = \frac{1}{2} \ln (x_a / x_b)$.
After integration over $k_{\perp a}, k_{\perp b}, x_a$, and $x_b$ the
result can be written in the form (the factor $1/q_\perp^{2\lambda}$
originates from the $x$-dependence of $\mathcal{F} \sim x^{-\lambda}$)
\begin{equation}
\frac{\rd\sigma_{\incl}^{\jet}}{\rd q_\perp^2 dy} \propto 
\frac{s^\lambda}{q_\perp^{4+2\lambda}} \cdot 
\alpha_s^2(q_\perp^2) \cdot h(q_\perp^2).
\label{our-res}
\end{equation} 
Each outgoing parton is connected to two links, and therefore counted
in {\em two} subcollisions in fig.~\ref{fig:fans}. Therefore, to avoid
{\em double counting}, we must include an extra factor $1/2$ in
eq.~(\ref{incljetdistr}). We can compare with the result of the
``naive'' approach in eq.~(\ref{naive}), and define a corresponding
function $h_{\naive}$. (We note that the functions $h(q_\perp^2)$ are
defined in such a way that scale independent
\begin{figure}[t]
\mbox{
\epsfxsize=24pc 
\epsfbox{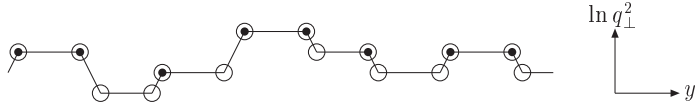}} 
\caption{A parton chain in the $(y,\ln q_\perp^2)$ plane. The emitted
partons are marked by dots. The ``naive'' approach corresponds to an
outgoing parton for each circle.  \label{fig:cirklar}}
\end{figure}
\begin{figure}[h]
\mbox{
\epsfxsize=13.5pc 
\epsfbox{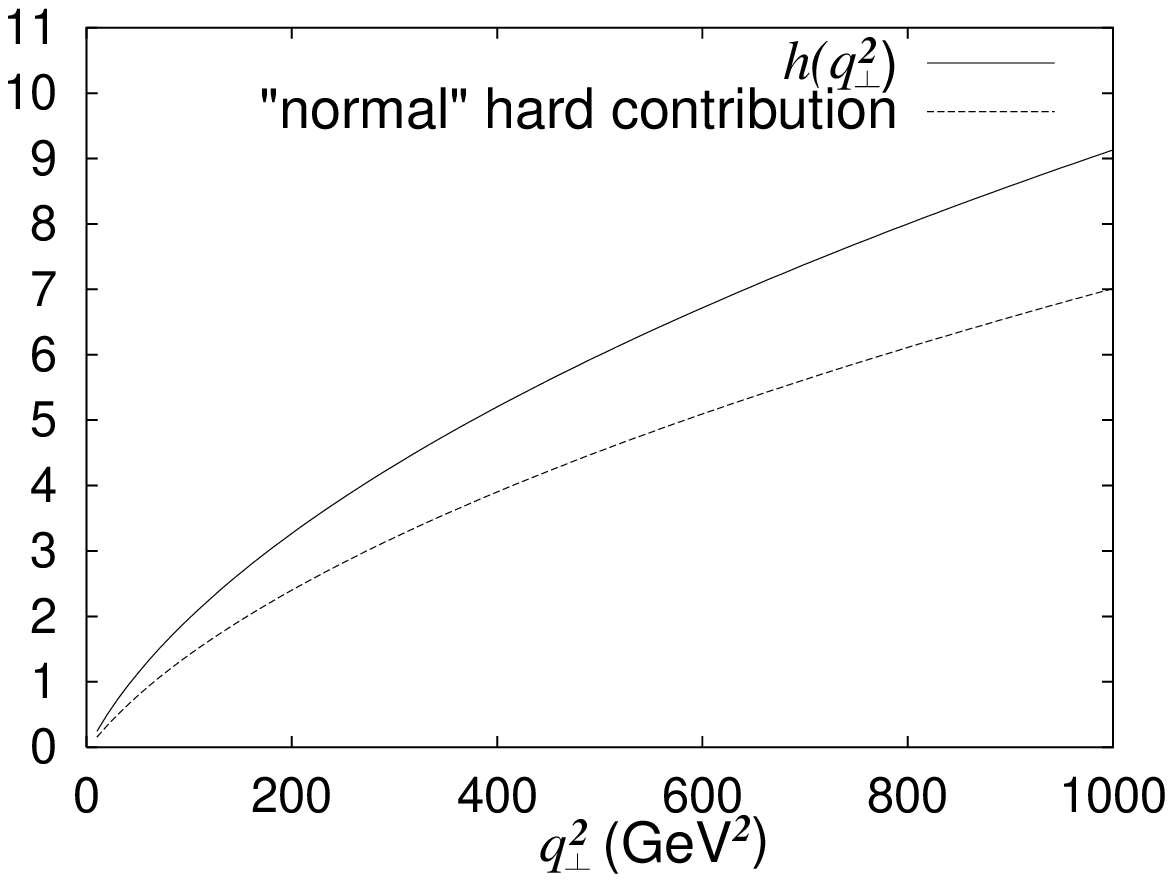}} 
\mbox{
\epsfxsize=13.5pc 
\epsfbox{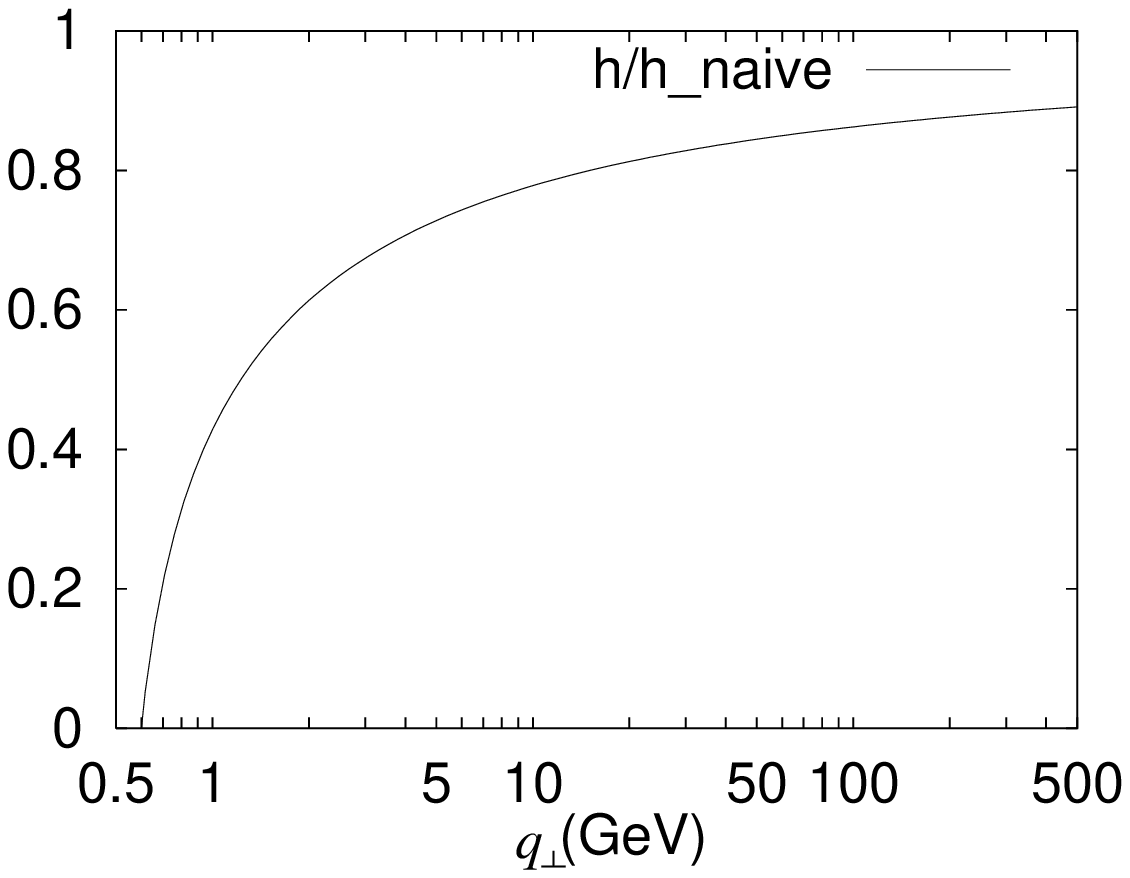}} 
\caption{The function $h(q_\perp^2)$ in eq.~(\ref{our-res}) (arbitrary
scale), and the ratio $h/h_{\naive}$.  \label{fig:h}}
\end{figure}
\begin{figure}[t]
\mbox{
\epsfxsize=13.5pc 
\epsfbox{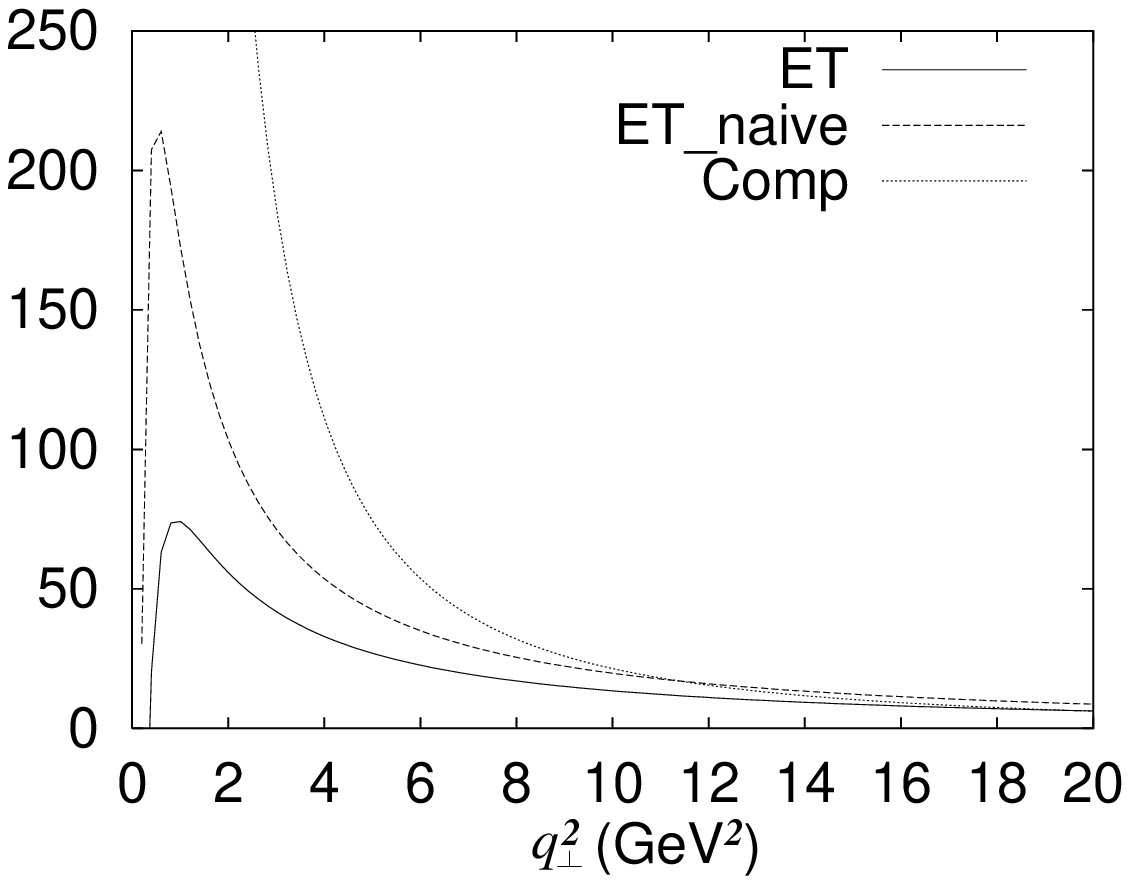}} 
\mbox{
\epsfxsize=13.5pc 
\epsfbox{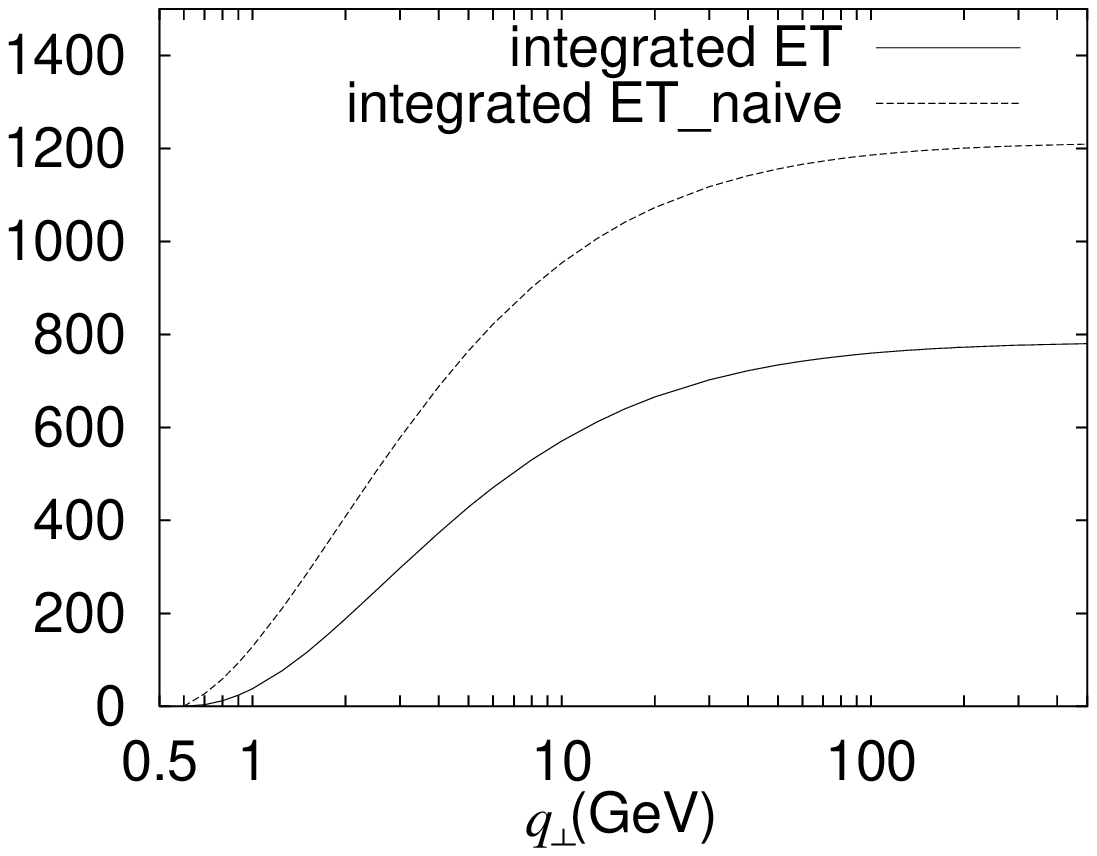}} 
\caption{ The $E_\perp$ distribution, $\rd E_\perp /\rd y \rd q_\perp^2$
(solid), compared to the ``naive'' estimate in eq.~(\ref{naive})
(dashed). For comparison we also show the result obtained for
scale-independent structure functions (dotted). Also shown is the
integrated $E_\perp$ distribution, $\int^{q_\perp^2} \rd q_\perp^2 \rd
E_\perp /\rd y \rd q_\perp^2$ (arbitrary scales).
\label{fig:Et}}
\end{figure}
 structure functions $F(x)$ in eq.~(\ref{naive}) would correspond to
 $h_{\naive}=constant$.) As discussed in more detail in \cite{miu} the
 expression in eq.~(\ref{naive}) corresponds to two jets for each link
 in the chain. This double counting does, however, not give a factor
 of 2 in the jet cross section, because the jets are expected to have
 transverse momenta given by the $k_\perp$ of the link, also when this
 is smaller than the $k_{\perp a}$ or $k_{\perp b}$ of the colliding
 partons. Therefore the distribution of {\em hard} jets becomes
 approximately correct, but the distribution of softer jets becomes
 strongly overestimated. This is illustrated in
 fig.~\ref{fig:cirklar}.

As seen in fig.~\ref{fig:h}, $h(q_\perp^2)$ is much reduced {\em
cf}. to $h_{\naive}$ for small $q_\perp$. This is also seen in the
$E_\perp$-flow presented in fig.~\ref{fig:Et}. From the integrated
$E_\perp$-distribution we see that the difference in total $E_\perp$
between our and the ``naive'' result corresponds roughly to the
$E_\perp$ flow below 2 GeV in the ``naive'' approach. Consequently
the total $E_\perp$ flow in our approach corresponds to the ``naive''
result, if the latter had a low $q_\perp$ cut-off around 2 GeV.
\section{Summary}
The Linked Dipole Chain model, developed for high energy DIS, is also
applicable to hard subcollisions in hh, hA or AA collisions. The
inclusive jet cross section is expressed in terms of {\em
non-integrated} structure functions ($k_\perp$-factorization). The
formalism also automatically {\em avoids double counting} for long
parton chains. The result gives a {\em dynamical suppression for small
$q_\perp$}, which corresponds to an effective cut-off around 2 GeV
in a ``naive'' approach in terms of integrated structure functions.
\section*{Acknowledgments}
The results presented in this talk are obtained in collaboration with
B. Andersson, J. Samuelsson, H. Kharraziha, L. L\"onnblad, and G. Miu.


\begin{thebibliography}{99}

\bibitem{cutoff}
T. Sj\"ostrand, Lund preprint LUTP 99-42, hep-ph/0001032.
  
\bibitem{LDC}
B. Andersson, G. Gustafson, J. Samuelsson, {\em Nucl. Phys.} 
B {\bf 467}, 443 (1996); B. Andersson, G. Gustafson, H. Kharraziha, 
{\em Phys. Rev.} D {\bf 57}, 5543 (1998).

\bibitem{LDCMC}
H. Kharraziha and L. L\"onnblad, Lund preprint, LUTP 97-34; 
{\em JHEP} {\bf 03}, 006 (1998).
  
\bibitem{CCFM}
M. Ciafaloni, {\rm Nucl. Phys.} B {\bf 296}, 49 (1988);
S. Catani, F. Fiorani, G. Marche-sini, {\em Phys. Lett.} B {\bf 234}, 339 
(1990);  {\em Nucl. Phys.} B {\bf 336}, 18 (1990).

\bibitem{HJLL}
H. Jung, L. L\"onnblad, {em J. Phys.} G {\bf 26}, 707 (2000).

\bibitem{miu}
G. Gustafson, G. Miu, {\em Phys.Rev.} D {\bf 63}, 034004 (2001).

\end{thebibliography}
\end{document}
